# Fission waves can oscillate

AG Osborne[a], MR Deinert[a]•

[a] Department of Mechanical Engineering, The Colorado School of Mines, Golden, CO 80401, USA.
• Corresponding author, phone: 303-384-2387, email: mdeinert@mines.edu

**Abstract**. Under the right conditions, self sustaining fission waves can form in fertile nuclear materials. These waves result from the transport and absorption of neutrons and the resulting production of fissile isotopes. When these fission, additional neutrons are produced and the chain reaction propagates until it is poisoned by the buildup of fission products. It is typically assumed that fission waves are soliton-like and self stabilizing. However, we show that in uranium, coupling of the neutron field to the $^{239}$U->$^{239}$Np->$^{239}$Pu decay chain can lead to a Hopf bifurcation. The fission reaction then ramps up and down, along with the wave velocity. The critical driver for the instability is a delay, caused by the half-life of $^{239}$U, between the time evolution of the neutron field and the production of $^{239}$Pu. This allows the $^{239}$Pu to accumulate and burn out in a self limiting oscillation that is characteristic of a Hopf bifurcation. Time dependent results are obtained using a numerical implementation of a reduced order reaction-diffusion model for a fast neutron field. Monte Carlo simulations in combination with a linear stability analysis are used to confirm the results for the full system and to establish the parameter space where the Hopf occurs.

**Introduction.** Reaction diffusion processes affect a remarkable range of phenomena, from neural communication and traffic flow, to the coloring of animal coats and the spread of epidemics (1-4). Under the right conditions, self stabilizing reaction waves can arise that will propagate at constant velocity. That such phenomena can arise in fertile nuclear materials dates to at least (5) and the subject gained considerable attention after Edward Teller's work on self propagating fission waves in thorium (6). While some reaction diffusion systems are known to exhibit instabilities (7), a common assumption with fission waves is that they are soliton-like (8), and asymptotically stable (9, 10). This has resulted in hopes that an inherently self regulating reactor could be designed on the basis of this phenomenon (5).

Fission waves are a form of breed-burn phenomena. The first mention of the concept in a nuclear context dates to at least 1958 and a meeting of the United Nations Atomic Energy Agency (11). The underlying mechanism is simple: a fertile material, typically uranium, would be subjected to a fast neutron field, with the neutrons transmuting the uranium to produce transuranics. If these built up to a sufficiently high concentration for criticality, more fast neutrons would result, thereby creating a self sustaining process that would propagate slowly. Fission products would build up until they poison the chain reaction in the tail of the propagating wave.

While hundreds of isotopes are present in any neutron chain reacting system (12), the overall behavior is largely determined by only a relatively small number of nuclides. The critical mechanism distinguishing a fission wave in uranium from a normal chain reaction is the capture of neutrons by $^{238}$U, and the subsequent self sustaining production and fission of higher actinides, Fig. 1. Simulations have shown that the velocity of a fission wave would be at most a few centimeters per year. Because of this, people have assumed that the half lives of $^{239}$U and $^{239}$Np are short relative to the rate of propagation, and in particular that $^{239}$U decays immediately to $^{239}$Np (13-15). From here, $^{239}$Pu and the transuranics are produced. This then results in a stabilizing feedback mechanism. If a perturbation causes the fission rate to increase, key transuranics are consumed faster than they are produced, and the fission rate goes back down. If the fission rate decreases, these transuranics are produced faster than they are consumed and the fission rate rises again.

In reality, $^{239}$U has a 23 minute half life and the neutron field couples to all of the transuranics, both short and long lived. This complicates the feedback between the neutron flux and the isotopics, and sets up the conditions under which a Hopf bifurcation can arise.

The dynamics of a fast neutron field can be well approximated using a one-group neutron diffusion equation (16):

$$\frac{\partial \phi}{\partial t} = v\left(\nabla \cdot D \nabla \phi + V\phi + L(z)\phi + P\right) + \gamma\phi \qquad (1).$$

Here $\phi$ is the neutron flux, $v$ is neutron velocity, $\gamma$ is a thermal reactivity term, $D$ is the neutron diffusion coefficient, $V$ is the difference in the production and consumption of neutrons at a point in the system, $L$ is the radial leakage of neutrons, and $P$ represents delayed neutrons produced during beta decay of fission products. The evolution of the isotopes in the system, can be described using a set of coupled first order reaction equations:

$$\frac{d\mathbf{N}}{dt} = (A\phi + B)\mathbf{N} \tag{2}$$

Here **N** is a column vector of nuclide concentrations, *A* is a square matrix representing the neutron reactions, and *B* is a square matrix of nuclear decay constants.

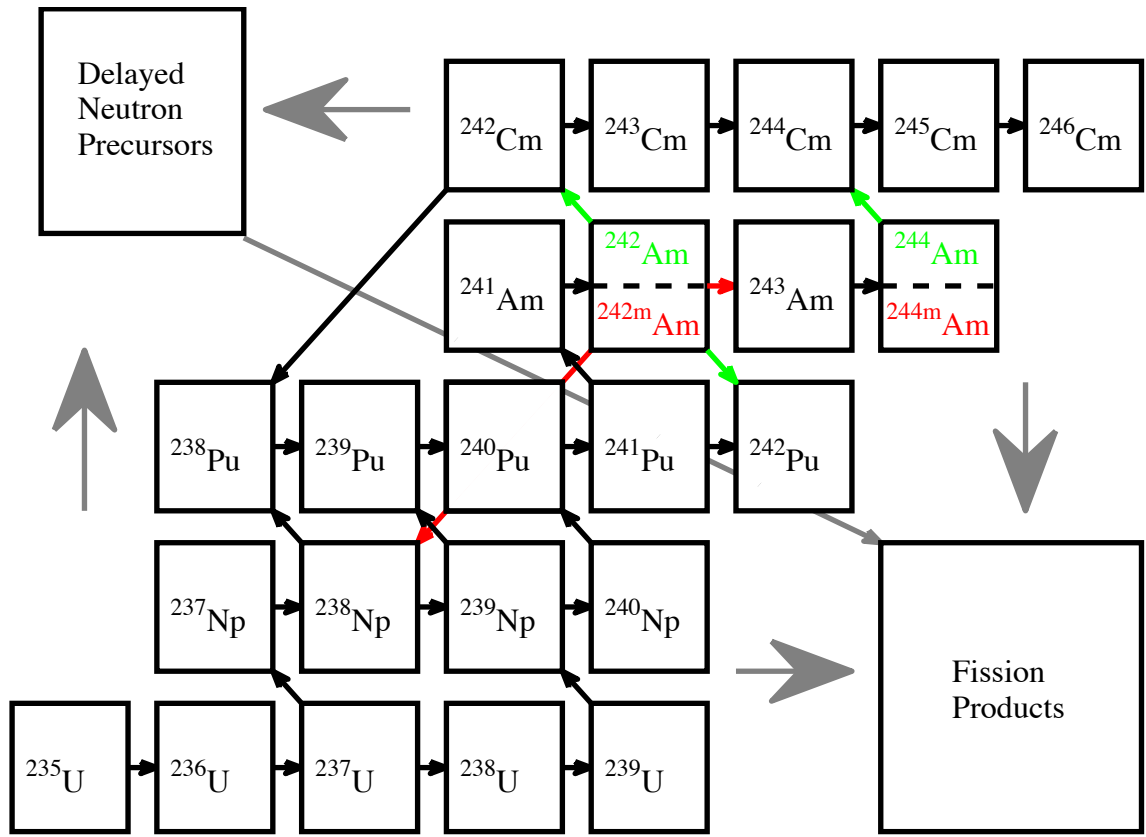

**Figure 1. Critical transuranics in a fission wave transmutation chain.** The direction of the arrows in the transmutation chain indicates reaction channel. Horizontal-right arrows represent neutron absorption, up-left arrows represent beta decay, down-left arrows represent alpha decays, down-right arrows represent electron capture or beta-plus decay. Large arrows represent fission in actinides, which produces fission products represented by a lumped pseudo-nuclide. A small proportion of fission reactions produce isotopes which experience delayed neutron-producing beta decays. These delayed neutron precursors are modeled as 6 groups of pseudo-nuclides, which feed into the lumped fission product following decay. $^{239}$U is central to the production of higher actinides. $^{239}$Pu is the dominant neutron producer in the fission wave, accounting for ~81% of neutron production.

For computational efficiency, the dependence of fission wave dynamics on system parameters were investigated in a reduced order system using a finite difference implementation of the 1-D form of Eq. (1). Here an implementation of Eq. (2) was used to track a reduced system of the dominant 38 nuclides that affect the kinetics, including a lumped fission product that encompassed the effects of 214 individual fission product isotopes. The simulations evolved to a point of instability, after which the fission wave entered into an oscillatory limit cycle, Fig. 2a. The instability and oscillation frequency of the wave were confirmed using a linear stability analysis of Eqs (1,2). A Taylor series was used to represent the time evolution of a perturbation to the system. The partial derivatives in the Taylor series were written as a Jacobian matrix, whose eigenvalues then identified the Hopf bifurcation (17), confirmed the oscillation frequency, and identified the driver of the bifurcation. In order to capture the collective behavior of the wave, the terms in the Jacobian were integrated over the positive $V$ region of the wave in a flux weighted manner. The results from the reduced order system were confirmed by repeating the linear stability analysis using the results from a 3-D, continuous energy spectrum Monte Carlo simulation. Here the fission wave was simulated with 1459 isotopes being tracked. The results were used to generate the inputs to the Jacobian, which were again integrated over the positive $V$ region of the wave in a flux weighted manner.

**Results**. The isotope $^{239}U$ was found to be the critical driver for the Hopf bifurcation shown in Fig. 2a. In the reference frame of the wave, its composition is fixed, with isotopes being produced at the same rate at which they are lost. In the absence of $^{239}U$ a positive perturbation to the flux increases both the rate at which $^{239}Pu$ is produced, and consumed. However, because of $^{239}Np$'s 2.3 day half life, the increase in the rate of production is smaller than the increase in the rate of consumption. The result is that the fission rate, and neutron flux, fall. This then decreases the rate at which $^{239}Pu$ is burned out, and this isotope builds back up as a result of the decay of $^{239}Np$. In addition, $^{239}Np$ couples to the neutron field and when this rises, the rate of its burnout increases. The burnout of $^{239}Pu$ and $^{239}Np$ act to limit the flux and damp the perturbation, Fig. 2b.

However, when $^{239}U$ is taken into consideration, its 23 minute half life causes the increase in the concentration of $^{239}Np$ to lag the flux. This allows the concentration of $^{239}Np$ to rise with time, and that of $^{239}Pu$ with it. This small phase shift is the origin of the Hopf bifurcation, Fig 2a. The dependence of the stability on $^{239}U$ is shown in Fig. 3 for both the reduced order and 3-D system with full isotopics. As the decay constant is artificially changed from 10 s$^{-1}$ through its natural value of ~4.9 x 10$^{-4}$ s$^{-1}$, the real part of the conjugate pair of eigenvalues crosses the imaginary axis on the phase plane, Fig. 3. The real parts of the conjugate pair of eigenvalues are -3.9×10$^{-7}$ and -6.3×10$^{-7}$ [1/s] for the reduced order and full systems for $T_{1/2}$=0, and converge to within 1% for the actual half life. The oscillation periods from the 1-D diffusion simulation and from the stability analyses for the reduced order and full systems were 5.38, 5.33 and 5.37 hours respectively, which agree to 1%.

The role of $^{239}U$ in the Hopf bifurcation was also shown by artificially halting the fission reaction, allowing the $^{239}U$ to decay away, and then allowing the fission wave to restart. As the $^{239}U$ builds back in, the Jacobian matrix for the reduced order system again shows a conjugate pair of eigenvalues whose real part crosses from negative to positive, Fig. 4. This shows how a Hopf bifurcation would arise in a physical system as the composition of the wave evolves, and $^{239}U$ builds in.

In physical neutron chain reacting systems thermal feedback can itself cause instability (18). In particular, in a sodium cooled fast reactor an increased fission rate drives up the temperature of the fuel and sodium coolant, causing them to expand. The lower coolant density hardens the neutron spectrum, which increases the fission-to-capture ratio, causing the fission rate to increase. Thermal feedback in fast reactors can be a strong effect, and from a practical point of view it is important to know whether this would overwhelm the dynamics that lead to the Hopf bifurcation.

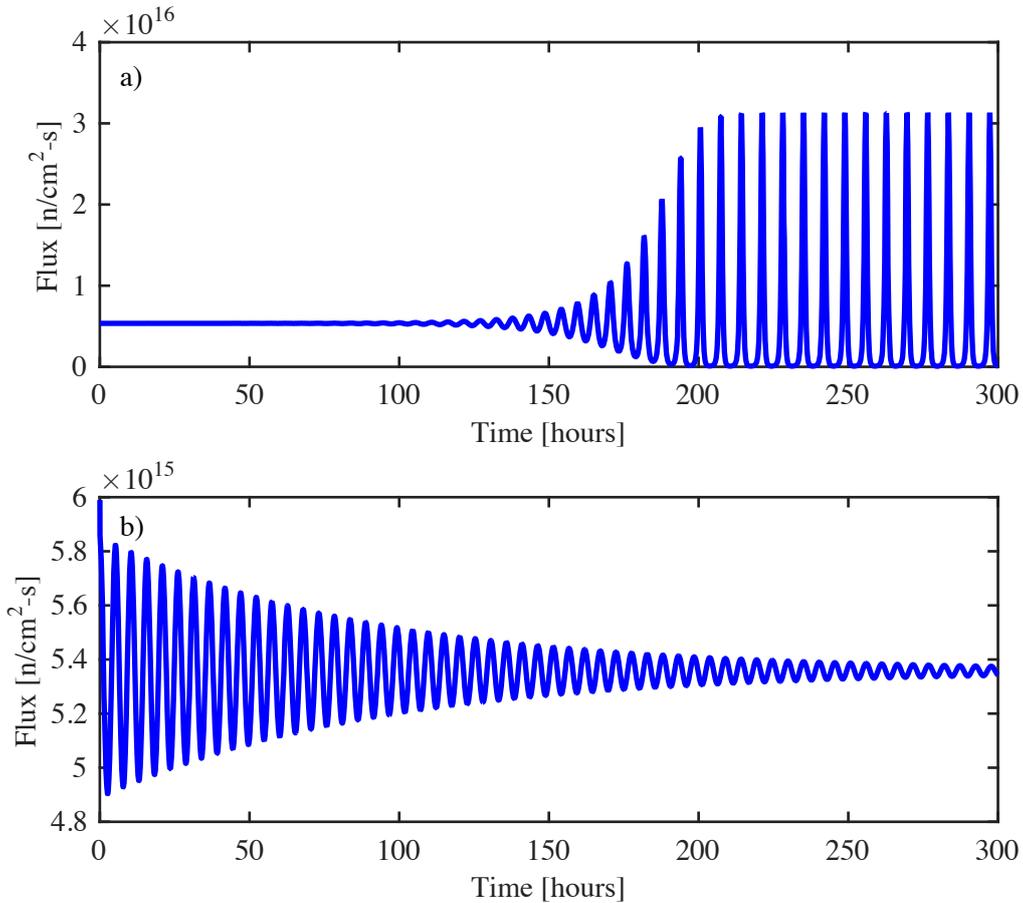

**Figure 2. Wave stability with and without $^{239}$U.** a) Perturbations to the neutron flux lead to a Hopf bifurcation when $^{239}$U is included in the transmutation chain. b) Perturbations to the flux damp out when $^{239}$U is not included in the transmutation chain.

To determine the effect of thermal feedback on the stability of the fission wave, the eigenvalues of the Jacobian representing the 3-D, 1459 nuclide system were recomputed as a function of the coolant and fuel reactivity coefficients $\alpha_C$ and $\alpha_F$. Figure 5 shows that there is a region in the $\alpha_C$ and $\alpha_F$ parameter space which would permit the Hopf bifurcation to arise in a metal fueled and sodium cooled reactor. Again, $^{239}$U is the critical parameter in the formation of the Hopf bifurcation. Figure 5 also shows the space of $\alpha_C$ and $\alpha_F$ values in which sodium cooled breed-burn reactor systems have been shown to lie (19-21). Importantly, this space has a significant overlap with the Hopf bifurcation region. However, the figure also shows that the de-stabilizing effect of isotopic feedback is negated by thermal feedback in some cases.

The results presented here are the first to show an isotopically driven mechanism for a Hopf bifurcation in a breed-burn system. This is also only the second example of an isotopically driven oscillation in a nuclear chain reacting system to be identified, the first being xenon oscillations in large thermal spectrum reactors (22). Importantly, the results indicate that the prevailing assumption of self-stability of fission waves is not true. The decay of $^{239}$U causes the concentration of $^{239}$Pu to increase with respect to changes in the neutron flux. This results in an oscillation that increases in amplitude until the rate of $^{239}$Pu production with each cycle equals the rate of burnout. Even in the presence of thermal feedback, an isotopic feedback could lead to the formation of a Hopf bifurcation.

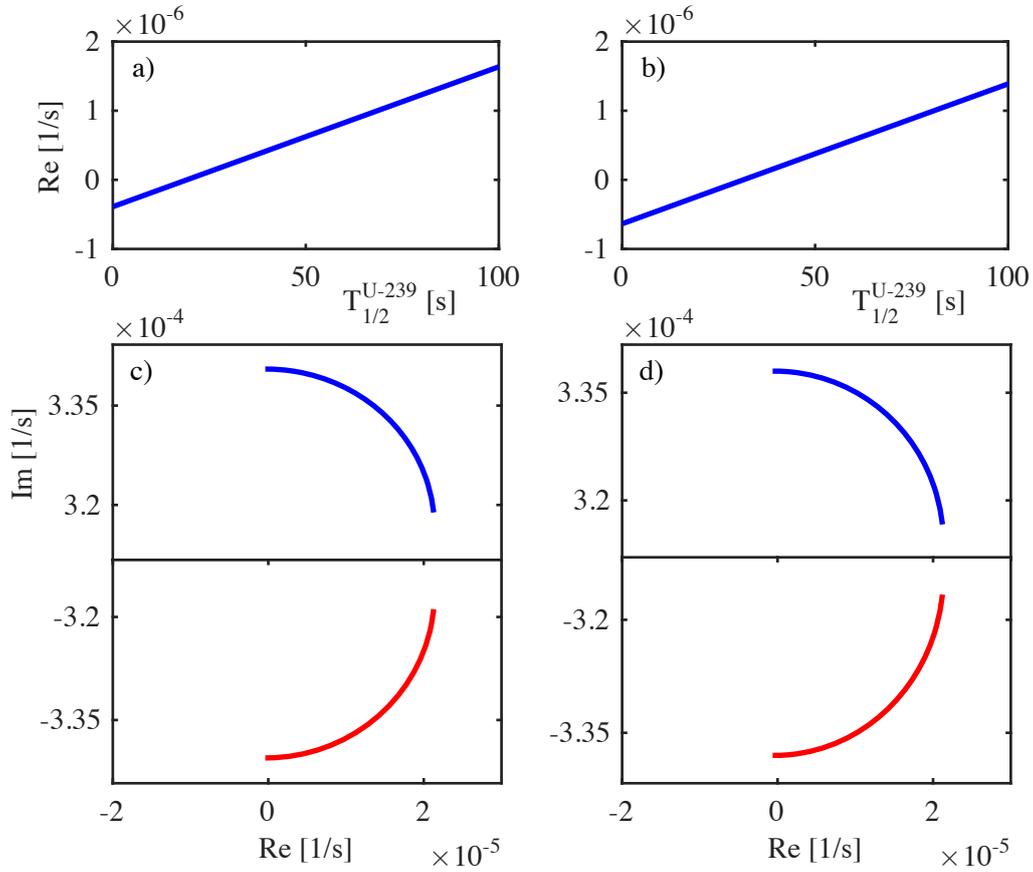

**Figure 3. Eigenvalues of Jacobian.** (top-left) Real part of the conjugate pair of eigenvalues as a function of the half-life of $^{239}$U for the reduced order system system, (top-right) real part of the conjugate pair of eigenvalues as a function of the half-life of $^{239}$U for the 3-D, 1459-nuclide system. In both cases the real part goes from negative to positive, identifying a Hopf bifurcation. (bottom-left) Trajectories of the conjugate pairs on the complex plane for the reduced order system as the half-life increases, (bottom-right) trajectories of the conjugate pairs on the complex plane for the 3-D, 1459-nuclide system as the half-life increases. Direction of trajectories are from left to right.

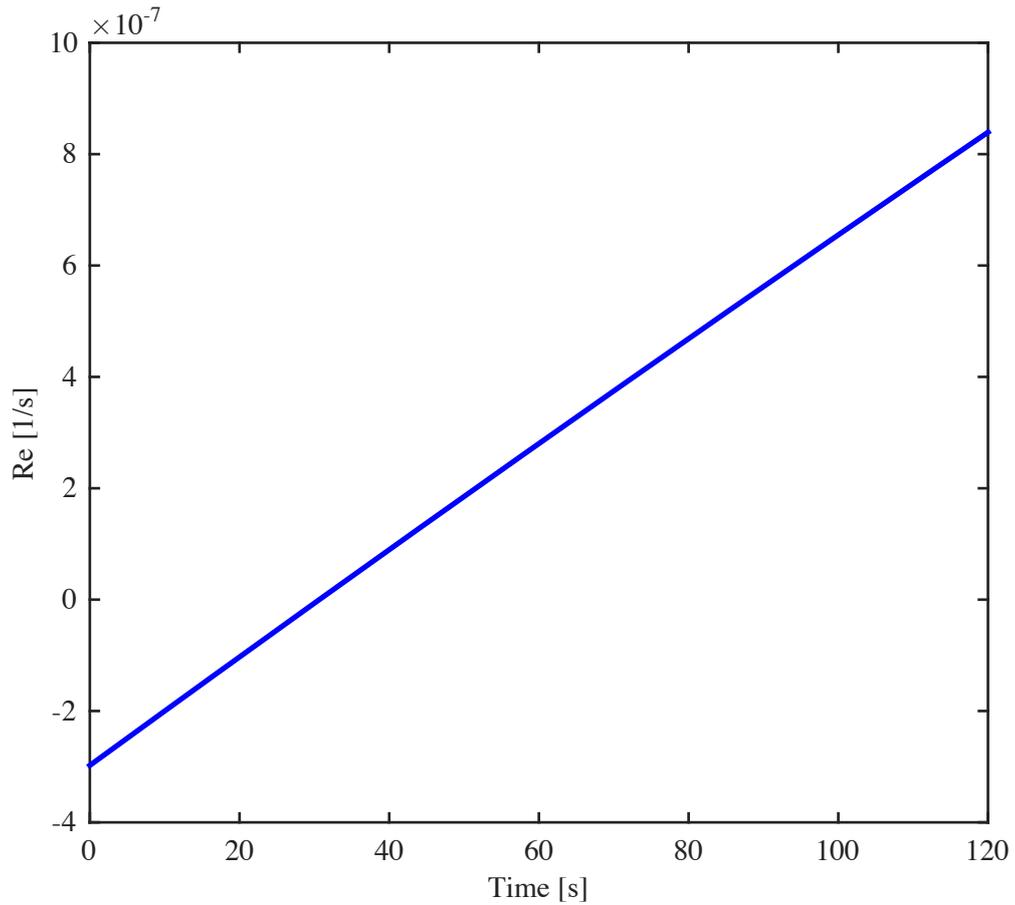

**Figure 4. Evolution of the conjugate pair with time as $^{239}$U builds in.** The figure shows the evolution of the real part of the conjugate pair of eigenvalues in the reduced order diffusion simulation. At t=0 the concentration of $^{239}$U is forced to zero, and as it builds back in from neutron capture in $^{238}$U the real part of the eigenvalue transitions from negative to positive, crossing zero after approximately 30 seconds.

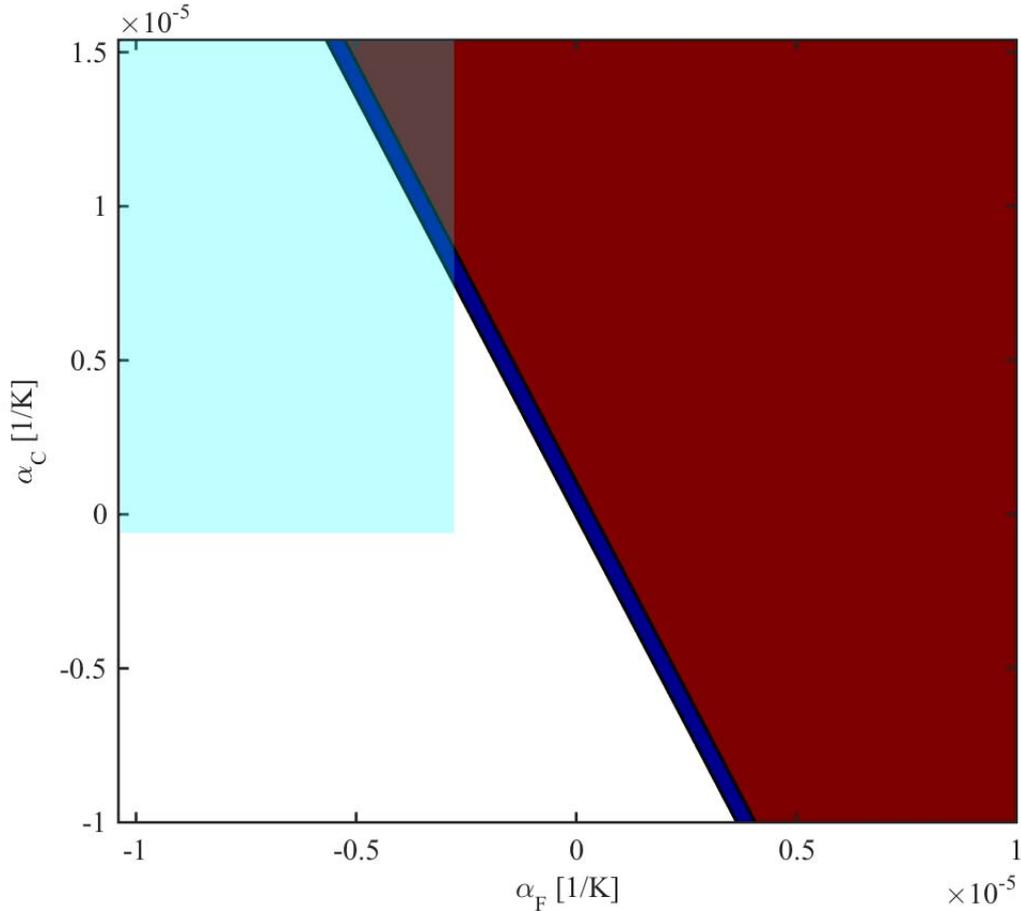

**Figure 5. Stability space with thermal feedback.** The stability space was evaluated for the 3-D, 1459-nuclide system by computing its eigenvalues while varying the coolant and fuel reactivity coefficients $\alpha_C$ and $\alpha_F$. The white and red regions indicate combinations of $\alpha_C$ and $\alpha_F$ that produce stable and unstable behavior, while the dark blue region shows the combinations of $\alpha_C$ and $\alpha_F$ that would give rise to a Hopf bifurcation. The light blue shaded region represents the range of reactivity coefficients that have been calculated in models of breed-burn systems. The figure shows that a Hopf bifurcation can arise even in the presence of the type of strong thermal feedback that occurs in real fast reactors.

**Methods**. A physical system with a structure typical of a fast spectrum reactor was used. The medium was initially composed of a homogeneous mixture of depleted uranium-zirconium fuel, liquid sodium coolant and structural steel, in the shape of a cylinder 10m in length and 160cm in radius. The 1-D form of equation (1) was solved numerically using a finite-difference discretization and a Newton-Raphson iteration step (23). During the iteration, Eq. (2) was solved using a matrix exponential solver. The use of Newton's method ensured that both the flux and the isotopics converged to machine precision at each time step. Inputs to Eqs. (1,2) were neutron cross sections, decay constants, scattering angles and prompt and delayed neutron multiplicity factors. Perturbations to the flux were used to induce oscillations.

The reactor was initially simulated using the Serpent 2.1.15 Monte Carlo code (24). This simulation was the source of the neutron flux data and cross section data for fission, capture and decay reactions. The remaining data were obtained from energy-dependent ENDF-B/VII.1 nuclear libraries (25). The libraries were processed into 494 energy bins using the NJOY software (26), then collapsed into a single energy bin by using the energy-dependent neutron flux from the Monte Carlo simulations as a weighting function. The Monte Carlo simulations tracked a total of 1459 nuclides. These include 214 fission product nuclides whose relative concentrations do not change significantly with burnup. These were modeled in the diffusion simulation by a stable lumped fission product whose cross sections represent the collective effect of the dominant 214 fission product isotopes.

The Monte Carlo simulation also tracked 236 nuclides that are unstable products of fission that emit a neutron during their decay into one of the 214 dominant fission products. These 236 nuclides were modeled by 6 groups of delayed neutron precursors whose overall behavior accurately reproduce the yield and decay characteristics of the 236 neutron emitting nuclides. Each of the 6 groups of delayed neutron precursors decays directly to the lumped fission product in the diffusion model. The Monte Carlo simulations also tracked 899 short-lived direct fission products that decay into the dominant 214 fission products without emitting a neutron. The half-lives of the 899 unstable nuclides, however, affect the timing of the overall system, and so in the diffusion simulation their aggregate behavior was also modeled using 6 groups of pseudo-nuclides that decay into the lumped fission product. Since 6 groups of pseudo-nuclides was adequate to model the delayed neutron precursors, we chose to use 6 groups of pseudo-nuclides to model the behavior of the 899 short-lived direct fission products as well.

The diffusion simulation tracked 25 actinides in the transmutation chain for $^{235}$U to $^{246}$Cm, Fig. 1. The 25 actinides were the ones that had the greatest effect on the dominant pair of eigenvalues of the 1459-nuclide system. The concentration of each actinide tracked by the Monte Carlo simulation was perturbed in sequence by an amount equal to $10^{-5}$ times its concentration at the stationary point, and the eigenvalues were recomputed. It was found that the total change in the imaginary part of the eigenvalues due to perturbing the 25 actinides of Fig. 1 represented more than 99.9% of the total change that was observed due to perturbing all of the actinides tracked by the Monte Carlo simulation.

A Jacobian matrix was constructed from the partial derivatives of Eqs. (1,2). The eigenvalues and eigenvectors of the Jacobian were computed for concentrations of $^{239}$U that increased with time, starting with a system containing no $^{239}$U. This system would represent a reactor that has been in a shutdown state for enough time such that the $^{239}$U has decayed away. The presence of a conjugate pair of eigenvalues with trajectories crossing the imaginary axis would identify a Hopf bifurcation, with the $^{239}$U concentration as bifurcation parameter. The trajectories of the conjugate pair as a function of the half-life of $^{239}$U were also computed in order to confirm that the $^{239}$U decay was responsible for the Hopf bifurcation.

The validity of the reduced order model was verified by performing the linear stability analysis using results from a 3-D Monte Carlo simulation that tracked 1459 isotopes. Inputs to the Jacobian were computed by sampling the nuclides and flux from the reaction-diffusion simulation and the 3-D 1459 nuclide Serpent simulation in the region where overall neutron production is greater than neutron capture. The average nuclide concentration and flux were computed in a flux-weighted fashion, and describe a stationary point of the system. The time evolution of the stationary point and the oscillation frequency were determined by eigenfunction expansion.

The overall change in reactivity due to changes in fuel temperature was given by:

$$\alpha_T = \alpha_F + \frac{\partial T_C}{\partial T_F} \alpha_C \tag{3}$$

where $\alpha_F$ and $\alpha_C$ are the fuel and coolant coefficients of temperature reactivity respectively [1/K]. $T_F$ and $T_C$ are the core averaged fuel and coolant temperatures respectively [K], and $\partial T_C/\partial T_F$ is the change in coolant temperature with respect to a change in the temperature of the fuel. A thermal resistor model (18) was used to compute fuel and coolant temperatures at every position in the 3-D, 1459 nuclide simulation. The power level was perturbed and the temperatures were recomputed. The $T_F$ and $T_C$ values were determined numerically by taking averages of the nominal and perturbed temperature profiles in the region where neutron production is greater than neutron capture, and computing their ratio. The stability regions shown in figure 5 were computed by varying $\alpha_F$ and $\alpha_C$ between $\pm 10^{-5}$ [1/K], and recomputing the eigenvalues at each pair of $\alpha_F$ and $\alpha_C$ values. The range of $\alpha_F$ and $\alpha_C$ values shown in the light blue shaded region represents values that breed-burn reactor models have been shown to exhibit (19-21). More detail can be found in.

The origin of the instability was verified by comparing the relative phase of the flux and $^{239}$Np production term when $^{239}$U is omitted in the transmutation chain, and when it is included. When $^{239}$U is omitted in the transmutation chain, peak flux values correspond to peak production rates of $^{239}$Np. Conversely, when $^{239}$U is included in the transmutation chain, the peak production rate of $^{239}$Np lags the peak flux slightly.

**Acknowledgments.** We thank Jaakko Leppanen and Tuomas Viitanen of the VTT Technical Research Centre of Finland for their helpful advice on the Serpent Monte Carlo code. We also thank the US Nuclear Regulatory Commission for NRC-38-08-946, which helped to support this work.